\documentclass[twocolumn,aps,prc,10pt,nofootinbib]{revtex4}

\usepackage[english]{babel} 	
\usepackage{graphics} 
\usepackage[pdftex]{graphicx}
\usepackage{amssymb} 			
\usepackage{amsmath} 			
\usepackage{booktabs}
\usepackage{tabularx}
\usepackage{rotating}

\usepackage{float}
\usepackage{url}	
\usepackage{bm}

\graphicspath{{./figures/}}

\begin{document}

\title{Few-body structures in the mirror nuclei, $^{11}$O and $^{11}$Li}

\author{E. Garrido$^{1}$, A.S. Jensen$^{2}$}

\affiliation{$^{1}$Instituto de Estructura de la Materia, IEM-CSIC, Serrano 123, E-28006 Madrid, Spain}

\affiliation{$^{2}$Department of Physics and Astronomy, Aarhus University, DK-8000 Aarhus C, Denmark} 

\date{\today}

\begin{abstract}
We investigate the dripline mirror nuclei, $^{11}$Li and $^{11}$O,
located on the neutron and proton dripline, respectively.  We calculate
the lowest four states, $3/2^-$, $1/2^+$, $3/2^+$ and $5/2^+$, built
on double occupancy in the nuclear $s_{1/2}$ and $p_{1/2}$ valence
single-particle states.  We use the hyperspherical adiabatic expansion
method to solve the three-body problem for a frozen nuclear core
surrounded by two identical nucleons.  The four analogue states in
$^{11}$O are obtained with precisely the same interactions as used for
the four states in $^{11}$Li, except for addition of the Coulomb
interaction from the charge of the substituted valence
protons. Surprisingly the four energies deviate from each other only
by less than a few hundred keV. Any of them could then turn out to be
the ground state, due to the uncertainty related to the angular
momentum and parity dependence of the three-body potential.  Still,
our calculations marginally favor the $1/2^+$ state.  The structures of
these four states in $^{11}$O deviate substantially from the analogue
states in the mirror, $^{11}$Li.
\end{abstract}


\maketitle


\section{Introduction}

Nuclear structure varies tremendously from the many-body leptodermous
features of heavy nuclei to the individual properties of light nuclei
\cite{boh75,zel17,sie87,mye69,tho04,fre07}.  The unexpected observed
increased jump in radius from lighter Li-isotopes to $^{11}$Li
\cite{tan85,tan85b} triggered the research on halo structures
\cite{han87} in a number of subfields of physics \cite{jen04}.
Few-body structure was especially efficient to describe the gross
features of halos, simply because the degrees of freedom essentially
decouple into two groups, where only a few nucleons determine the
low-energy properties \cite{jen04}.

The overall properties of $^{11}$Li are established as a three-body
system with constituents of two neutrons and $^{9}$Li \cite{joh90}.
In this connection the spin-spin splitting of the $s_{1/2}$ and
$p_{1/2}$ single-neutron states coupled to the $3/2^-$ ground state of
$^{9}$Li is crucial for the halo properties \cite{kat99,gar02a}.
These halo structures  are consistent with reaction information
\cite{gar96,gar97,cas17} even after the binding energy has been
measured with better accuracy \cite{smi08}.  The space spanned by
these single-particle states provides the $3/2^-$ ground-state as well
as the dipole-excited states of $1/2^+$, $3/2^+$ and $5/2^+$
\cite{gar02}.

The present investigation is triggered by the recent experiments
\cite{web19} on the mirror nucleus, $^{11}$O, which was preceded by
related experiments on $^{10}$N \cite{hoo17,lep02}, and quickly
followed up by theoretical papers on these and a few neighboring
nuclei \cite{wan19,mor19,for19}.  Several previous publications on
$^{11}$O and $^{10}$N are available \cite{aoy97,til04,cha12}.  The
comparison of isobaric analogue structures is a classical nuclear
discipline, which has provided strong support for the generalization
of the isospin concept from nucleons to nuclei
\cite{boh75,zel17,sie87}.  Since halo structures only occur near
threshold for $s$ and $p$ valence-nucleons, as in $^{11}$Li, the
structure may be strongly influenced even for small energy changes.
Thus, mirror nuclei on the driplines are most likely to exhibit larger
differences than stable nuclear mirrors \cite{jen04}.

The mirror pair, $^{11}$Li and $^{11}$O, are located on the neutron
dripline and slightly outside the proton dripline,
respectively. Still, both are accessible by experiments, which for
$^{11}$O largely is possible due to the Coulomb barrier.  The major
effect is from the Coulomb interaction of the additional protons,
which has both direct and indirect influence.  The same structure in
both nuclei produces an energy difference from the additional charge.
However, the structure itself is modified by this extra Coulomb
interaction and in turn resulting in a modified energy.

The total effect of the additional Coulomb interaction is quantified
in the Thomas-Ehrman shift \cite{ehr51,tho52}, which is defined as the
energy difference (apart from the neutron-proton mass difference)
between analogue states in mirror nuclei.  This energy shift may
depend on the state, and probably therefore is especially sensitive
towards variation of halo structure between analogue states in
$^{11}$Li and $^{11}$O \cite{aue00,gri02,gar04}.  The possible
structure variation between these analogue states may lead to sizable
state-dependent Thomas-Ehrman shifts.  This may even change the
sequence of ground and excited states built on these valence
configurations.

The recent experimental activity towards $^{11}$O and $^{10}$N
is an opportunity to compare properties in mirrors each located around
different nucleon driplines. This has earlier proved to be
informative.  Previous theoretical investigations already provided a
number of details on these nuclei.  However, they are a little random
as essentially all are incomplete in descriptions of the low-lying
states supported by the valence nucleon $s_{1/2}$ and $p_{1/2}$
single-particle states, since only the $3/2^-$-state is considered.
The only exception is Ref.\cite{wan19}, where the positive parity
state $5/2^+$ is also investigated.

In general, the connection between these mirror nuclei is not
particularly well explored.  Furthermore, the previous results are for
some reason quantitatively deviating, either due to different methods,
interactions, or perhaps accuracy of some kind.  We therefore decided
to investigate these low-lying nuclear states by use of our well
established few-body method, which due to the phenomenological input
also is both simple and accurate.  Thus, in the present paper we
report on detailed studies of low-energy properties of $^{11}$O in
comparison to similar investigations of $^{11}$Li.  Our purpose is
two-fold, that is first to discuss few-body properties of the specific
$^{11}$O-nucleus, and second to look for general conclusions by 
studying this mirror of the prototype of a halo nucleus.  Larger
differences can be expected for such dripline structures in comparison
and in contrast to stable mirror nuclei.

The paper, in section II, first briefly presents the applied
hyperspherical adiabatic expansion method \cite{nie01}, the
degrees-of-freedom, and the choice of interaction form.  Section III
describes the choice of parameters and the derived properties of the
subsystems, $^{10}$Li and $^{10}$N.  Section IV, V and VI are devoted
to the computed three-body properties of $^{11}$O specifically in
comparison to $^{11}$Li.  In section VII we present a summary and the
conclusion.

\section{Sketch of the method}
\label{method}

The three-body calculations will be performed using the well-established
hyperspherical adiabatic expansion method described in detail in \cite{nie01}.
In this method the three-body wave function, with total angular momentum $J$
and projection $M$, is written as:
\begin{equation}
\Psi^{JM}=\frac{1}{\rho^{5/2}} \sum_n f^J_n(\rho) \Phi^{JM}_n(\rho,\Omega),
\label{exp0}
\end{equation}
where $\rho$ is the hyperradius, $\Omega$ collects the five
hyperangles as defined for instance in \cite{nie01}, and $f^J_n(\rho)$
are the radial expansion functions.  The basis set
$\{\Phi^{JM}_n(\rho,\Omega)\}$ used in the expansion above is formed
by the eigenfunctions of the angular part of the Schr\"{o}dinger (or
Faddeev) equations,
\begin{equation}
\left[ \hat{\Lambda}^2 + \frac{2m\rho^2}{\hbar^2}(V_{12}+V_{13}+V_{23}) \right] \Phi_n^{JM}=\lambda_n(\rho) \Phi_n^{JM}(\rho,\Omega),
\label{angf}
\end{equation}
where $\hat{\Lambda}$ is the grand-angular momentum operator \cite{nie01},
$V_{ij}$ is the interaction between particles $i$ and $j$, 
$m$ is the normalization mass used to define the Jacobi coordinates \cite{nie01}, 
and $\lambda_n(\rho)$ is the eigenvalue associated to the angular eigenfunction
$\Phi_n^{JM}(\rho,\Omega)$.

In practice, Eq.(\ref{angf}) is solved after the expansion
\begin{equation}
\Phi^{JM}_n(\rho,\Omega)=\sum_q  C_q^{(n)}(\rho) 
\left[ {\cal Y}_{\ell_x\ell_y}^{K L}(\Omega) \otimes \chi_{s_x s_y}^S \right]^{JM},
\label{exp1}
\end{equation}
where $q$  collects all the quantum
numbers $\{K,\ell_x,\ell_y,L,s_x,S\}$, where $\ell_x$ and $\ell_y$ are the 
relative orbital angular momenta between two of the particles, and between the
third particle and the center-of-mass of the first two, respectively. The total
orbital angular momentum $L$ results from the coupling of $\ell_x$ and $\ell_y$. 
The quantum number $K$ is the so-called hypermomentum, which is defined as 
$K=2\nu +\ell_x +\ell_y$, with $\nu=0,1,2,\cdots$. The dependence on these quantum numbers,
$\ell_x$, $\ell_y$, $L$, and $K$, is contained in the usual hyperspherical 
harmonics, ${\cal Y}_{\ell_x\ell_y}^{K L}(\Omega)$, whose definition can also
be found in \cite{nie01}, and which satisfy $\hat{\Lambda}^2 {\cal Y}_{\ell_x\ell_y}^{K L}=
K(K+4){\cal Y}_{\ell_x\ell_y}^{K L}$.
In the same way, $s_x$ is the total spin of two of the particles, which couples to the 
spin of the third particle, $s_y$, to give the total spin $S$. The total spin function
is represented in Eq.(\ref{exp1}) by $\chi_{s_x s_y}^S$. Finally $L$ and $S$ couple to the
total three-body angular momentum $J$ with projection $M$. 

Obviously the definition of the $\bm{x}$ and $\bm{y}$ coordinates (the Jacobi coordinates) is not unique, since
for three-body systems  three different sets of Jacobi coordinates can be formed \cite{nie01}. When
solving the Schr\"{o}dinger equation a choice has to be made, which means that only one
of the internal two-body subsystems is treated in its natural coordinate. In this work, however,
we solve instead the Faddeev equations, which have the nice property of treating all the three possible sets
of Jacobi coordinates on the same footing \cite{nie01}.

The radial functions, $f^J_n(\rho)$, in Eq.(\ref{exp0}) are
obtained after solving the set of coupled equations
\begin{eqnarray}
\lefteqn{
\hspace*{-1cm}
\left[  
-\frac{\partial^2}{\partial \rho^2}+\frac{\lambda_n(\rho)+\frac{15}{4}}{\rho^2} -\frac{2mE}{\hbar^2}
\right] f_n^J( \rho)=
} \nonumber \\ & &
\sum_{n'} \left(2P_{nn'}(\rho)\frac{\partial}{\partial\rho} +Q_{nn'}(\rho) \right) f_{n'}^J(\rho),
\label{radf}
\end{eqnarray}
where $E$ is the three-body energy, and  $\lambda_n(\rho)$ is obtained from the angular equation (\ref{angf}). 
The explicit form and properties of the coupling functions $P_{nn'}(\rho)$ and 
$Q_{nn'}(\rho)$ can be found in \cite{nie01}.

The set of equations (\ref{radf}) has to be solved imposing to the
radial wave functions the appropriate asymptotic behaviour. This is
particularly simple for bound states, due to the asymptotic
exponential fall-off of the radial wave functions. In order to exploit
the simplicity of this asymptotic behaviour, we compute resonances
(understood as poles of the $S$-matrix) by means of the complex
scaling method \cite{ho83,moi98}.  In this method the three-body
energy is allowed to be complex, and the radial coordinates are
rotated into the complex plane by an arbitrary angle $\theta$ ($\rho
\rightarrow \rho e^{i\theta}$). Under this transformation, and
provided that $\theta$ is sufficiently large, the resonance wave
function behaves asymptotically as a bound state, i.e., it decays
exponentially at large distances, and its complex energy,
$E=E_R-i\Gamma_R/2$, gives the resonance energy, $E_R$, and the
resonance width, $\Gamma_R$.

Being more specific, after the complex scaling transformation,
the Eqs.(\ref{radf}) are solved by imposing a box boundary condition. The continuum spectrum is then discretized, and
the corresponding discrete energies appear in the complex energy plane rotated by an angle equal to $2\theta$ \cite{ho83,moi98}.
The resonances show up as discrete points, independent of the complex scaling angle, and out of the  cut corresponding to continuum states.

Note that an accurate enough solution of the three-body problem requires convergence at two
different levels. First, one needs convergence in the expansion of the angular eigenfunctions in Eq.(\ref{exp1}), which 
is necessary in order to obtain sufficient accuracy in the $\lambda_n$-eigenvalues in the radial equations
(\ref{radf}). A correct convergence 
requires inclusion of the relevant $\{\ell_x,\ell_y,L,s_x,S\}$-components, and, for each of them, a sufficiently
large value, $K_{max}$, of the hypermomentum $K$ is also needed. Second, one has to reach convergence as well in the 
expansion in Eq.(\ref{exp0}), which implies a sufficiently large number of adiabatic terms. 

Typically, the 
convergence in the expansion (\ref{exp0}) is rather fast, and for bound states and resonances (after the 
complex scaling transformation) four or five
terms are usually enough. However, the expansion (\ref{exp1}) is more demanding, especially when 
dealing with particles with non-zero spin, since the number of components
can increase significantly in accordance with a given total three-body angular momentum $J$. Also, for extended systems, for which
the $\lambda_n$-functions have to be accurately computed at large distances, the required maximum
value of the hypermomentum, $K_{max}$, can be rather large.

Given a three-body system, the key quantities determining its properties are the two-body potentials entering in Eq.(\ref{angf}). In this work we shall assume that the nucleon-nucleon interaction is the GPT potential described in \cite{gog70}.

For the core-nucleon potential we choose an interaction, adjusted independently for the different partial waves, each term of the form:
\begin{equation}
V_{Nc}^{(\ell)}(r)=V_c^{(\ell)}(r)+V_{ss}^{(\ell)}(r) \bm{s}_c\cdot (\bm{\ell}+\bm{s}_N)+V_{so}^{(\ell)} \bm{\ell}\cdot \bm{s}_N,
\label{eq1}
\end{equation}
where $\bm{\ell}$ is the relative orbital angular momentum between the
core and the nucleon, whose intrinsic spins are denoted by $\bm{s}_c$
and $\bm{s}_N$, respectively. As shown in \cite{gar03}, this
spin-operator structure, which is consistent with the mean-field
description of the nucleons in the core, is crucial for a correct
implementation of the Pauli principle.

Obviously, when the interaction involves two charged particles, the Coulomb potential should be added to the interactions
described above. In this work  we shall describe the core as a uniformly charged sphere with radius equal to the charge radius,
which for  $^9$C will be taken equal to 2.5 fm.   We assume all nucleons are point-like particles.

\section{The core-nucleon system}
\label{pots}

For the case of $^{11}$Li ($^9$Li+$n$+$n$) and its mirror partner,
$^{11}$O ($^9$C+$p$+$p$), it is clear that the essential ingredient is
the nuclear part of the core-nucleon interaction.  Due to the charge
symmetry of the strong interaction, these potentials will be the same
for both, $^{10}$Li ($^9$Li+$n$) and $^{10}$N ($^9$C+$p$), since also
the $^{9}$Li and $^{9}$C cores are mirror nuclei.  Table~\ref{tab1a}
contains the parameters used in this work for the potential form given
in Eq.(\ref{eq1}) with the $s$- and $p$-state parameters from
Ref.\cite{mor19}.  The radial shapes are for convenience chosen to be
Gaussians with the same range in all terms.  The actual shape is
unimportant as long as it is of short range with a range consistent
with the core-size.

\begin{table}
  \begin{tabular}{|c|c|c|c|}
    \hline    
  $\ell$ &  $S_{c}^{(\ell)}$  &  $S_{ss}^{(\ell)}$  & $S_{so}^{(\ell)}$  \\ \hline
  0 &$-5.4$   & $-4.5$ & --    \\ 
  1 & 260.75 & 1.0  & 300    \\ 
  2 &  260   & $-9.0$ & $-300$    \\ \hline    
   \end{tabular}
  \caption{The strength parameters, $S_i^{(\ell)}$, in MeV for the
    Gaussian core-nucleon potentials, $V_i^{(\ell)}=S_i^{(\ell)}
    e^{-r^2/b^2}$, defined in Eq.(\ref{eq1}), with the $s$ and $p$
    partial waves as in Ref.\cite{mor19} (also denoted P1I in Ref.\cite{cas17}).  We
    choose the same numerical value, $b = 2.55$~fm, for the range
    parameter, $b$, in all terms and partial waves.  }
\label{tab1a}
\end{table}

The two all-decisive properties of the nucleon-core system are the positions
of the two-body resonances, and the exclusion of Pauli forbidden states
occupied by the core-nucleons.  The first property is achieved by the
numerical values specified in Table~\ref{tab1a}.
The second property is fulfilled by use of the shallow $s$-wave
potential without a bound state, and a large and inverse (positive)
sign of the $p$-wave spin-orbit strength, which places the
$p_{3/2}$-shell at an unreachable high energy.  In this way, by
construction, the valence-nucleon can not occupy the Pauli forbidden
$s_{1/2}$- and $p_{3/2}$-shells, which already are occupied by the six
neutrons or the six protons in the $^9$Li or $^9$C-core.

\begin{figure}
\begin{center}
\includegraphics[width=\linewidth,angle=0]{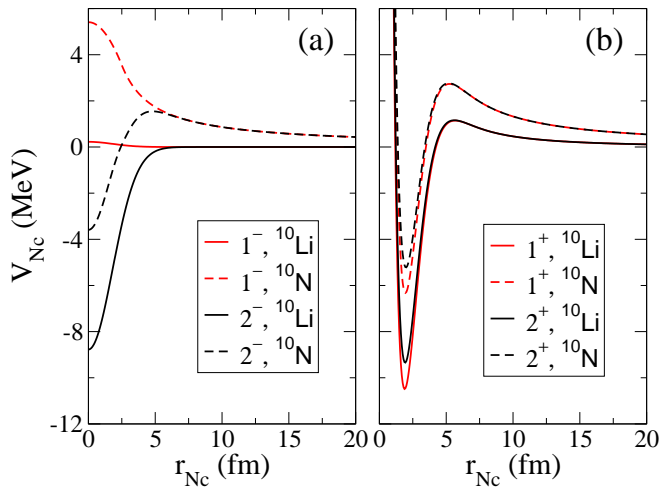}
\end{center}
\caption{(a): Core-nucleon potentials for $s_{1/2}$ states, $1^-$ (red) and $2^-$ (black), in $^{10}$Li (solid lines) and $^{10}$N (dashed lines). (b): The 
same as in panel (a), but for the $p_{1/2}$ states, $1^+$ (red) and $2^+$ (black).}
\label{figpot}
\end{figure}

The resulting nucleon-core potentials are shown in Fig.~\ref{figpot}
for the $s_{1/2}$-states, $1^-$ (red) and $2^-$ (black) in panel (a),
and for the $p_{1/2}$-states, $1^+$ (red) and $2^+$ (black) in panel
(b).  The solid and dashed lines refer, respectively, to the $^{10}$Li
and $^{10}$N-cases.  The difference between them arises from the 
Coulomb repulsion entering in the $^ 9$C-proton interaction for $^{10}$N. 
The $s$-waves in the left panel of
Fig.~\ref{figpot} reveal our choices for $^{10}$Li of an attractive
$2^-$-potential placing a virtual nucleon-core state very close to
zero energy, while in contrast the $1^-$-potential is very small and
slightly repulsive.  The same potentials for $^{10}$N are pushed up by
the Coulomb repulsion, where the $2^-$-potential still has an
attractive short-range part, whereas the $1^-$-potential is clearly
overall repulsive.  The $p$-wave potentials in the right panel of
Fig.~\ref{figpot} all have an attractive short-range part leading to
more or less known $p$-wave resonances in both $^{10}$Li and $^{10}$N.

As shown in \cite{gar02a,gar97}, the main properties of $^{11}$Li, as well 
as the behavior of the momentum distributions, are essentially determined
by the energy of the centroid of the spin-splitted $s$- and $p$- doublets. 
Therefore, the subsequent three-body results would remain basically 
unchanged with the opposite order of the $1^-$ and $2^-$-virtual states
and of the $1^+$ and $2^+$-resonances.

\subsection{$^{10}$Li-properties}

\begin{table}
\begin{footnotesize}
\begin{center}
\begin{tabular}{|c|cc|cc|cc|c|}
\hline    
      &\multicolumn{2}{c|}{This work} &\multicolumn{2}{c|}{Ref.\cite{kat99}} &  \multicolumn{2}{c|}{Exp.\cite{boh93}} & $\delta(E_R)=\frac{\pi}{2}$ \\ \hline
       &  $E_R$ &  $\Gamma_R$  &  $E_R$  &  $\Gamma_R$  &  $E_R$  &  $\Gamma_R$ &  $E_R$\\ \hline
       1$^-$ &   --    &  --    &   --  & --   &  --  & -- & --  \\
       2$^-$ &  $-0.020$ &  --             &  $-0.028$   &  --  &  --  &  -- & -- \\
       1$^+$ &  0.32 & 0.19             &  0.42  &  0.19  &    $0.42\pm0.05$ & $0.15\pm0.07$ & 0.37 \\
       2$^+$ &  0.58&  0.49              & 0.71   &  0.40 &     $0.80\pm0.08$   &  $0.30\pm0.10$ & 0.78 \\
       4$^-$ &  3.95&  2.45              &   4.13     &   3.12    &  $4.47\pm0.10$  & $0.7\pm0.2$ & 4.88 \\  \hline    
\end{tabular}
\end{center}
\end{footnotesize}
\caption{For $^{10}$Li, the second column shows the energies of the 1$^-$ and 2$^-$ virtual states, and energies and widths of the $1^+$, $2^+$ and $4^-$ resonances in $^{10}$Li 
obtained with the two-body potentials described in the text. The third column shows the energies and widths obtained in Ref.\cite{kat99}. In the fourth column
the available experimental data are given \cite{boh93}. The last column gives the resonance energies computed as $\delta(E_R)=\pi/2$. All the  energies, $E_R$,  and 
widths, $\Gamma_R$, are given in MeV.}
\label{tab10Li}
\end{table}

The $^{10}$Li-properties are determined by the potentials given by the solid curves in Figs.~\ref{figpot}a and \ref{figpot}b.
The computed spectrum is shown in the second column of Table~\ref{tab10Li}. The  ground state is a virtual $2^-$-state resulting
from the coupling of an $s_{1/2}$ valence-neutron with the $3/2^-$ ground-state of the core, whose energy is about $-20$ keV. 
The corresponding potential is given by the solid black curve in Fig.~\ref{figpot}a. Due to the repulsive character of the potential
shown by the solid red curve in the same figure, the $1^-$ $s$-wave partner appears at high energy in the continuum. The 
$p$-wave resonant-states, 1$^+$ and $2^+$, produced by the $p$-wave potential barriers (solid curves in Fig.~\ref{figpot}b), are 
found at 0.32 MeV and 0.58 MeV, respectively, with corresponding widths of 0.19 MeV and 0.49 MeV.

\begin{figure}
\begin{center}
\includegraphics[width=\linewidth,angle=0]{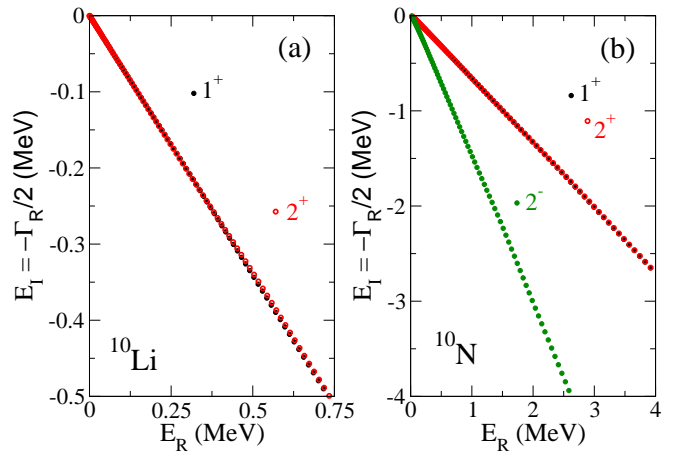}
\end{center}
\caption{Complex energies of the $p_{1/2}$-resonances in $^{10}$Li, panel (a), and the $s_{1/2}$- and $p_{1/2}$-resonances in $^{10}$N, panel (b),  after a 
two-body complex scaling calculation using the potential described in  Sect.~\ref{pots}.}
\label{fig1}
\end{figure}

The virtual state is obtained by finding the energy providing the correct divergent asymptotic behaviour produced by the
poles of the $S$-matrix located on the negative imaginary axis in the complex momentum plane. The resonances are also obtained as 
poles of the $S$-matrix by means of the complex scaling method \cite{ho83,moi98}, which simplifies the numerical calculation by giving rise to an exponential fall-off of the complex rotated resonance wave functions. As mentioned in Sect.~\ref{method}, the complex rotated two-body problem is solved after discretization of the continuum by means of a box boundary condition.

The corresponding discrete energies appear in the complex energy plane
rotated by an angle equal to twice the angle used for the complex
scaling coordinate transformation. The resonances appear as discrete points,
independent of the complex scaling angle, out of the cut (lines)
corresponding to continuum states. This is shown for $^{10}$Li in
Fig.~\ref{fig1}a. As we can see, a complex scaling angle of
$\theta=0.3$ rads is enough to capture the $1^+$ and
$2^+$-resonances. As shown in Table~\ref{tab10Li}, together with the
$1^+$ and $2^+$-states, the core-neutron potential described in
Sect.~\ref{pots} gives rise to a $4^-$-resonance (with the valence
neutron in the $d_{5/2}$-state) at 3.95 MeV with a width of 2.45 MeV.
For the sake of clarity in the figure, this resonance is not shown in
Fig.~\ref{fig1}a.

The computed spectrum can be compared to the one obtained in
Ref.\cite{kat99} (third column of Table~\ref{tab10Li}), where a
microscopic coupled-channel calculation is performed.  The similar
virtual $s$-states in the two calculations are dictated by a demand to
reproduce measured two-neutron halo properties of $^{11}$Li in
subsequent calculations. In the fourth column of the table we give the
available experimental data. Note that the ones of the $4^-$-state are
actually in Ref.\cite{boh93} assigned preliminary  to angular momentum
and parity, $2^-$. However, as suggested in \cite{kat99}, the
calculations might indicate that they could actually correspond to the
$4^-$-resonance.

Although in both this work and Ref.\cite{kat99}, the energies are
computed as poles of the $S$-matrix, the agreement with the
experimental $1^+$ and $2^+$-energies \cite{boh93} seems to be worse 
in our calculation. The $p$-states
deviate somewhat by more than about 100~keV in centroid energy of the two spin-split
$p$-states (0.5~MeV in this work and 0.6~MeV in \cite{kat99}) . 
However, in \cite{kat99} the calculation is performed by
fitting the energy of the $1^+$$S$-matrix pole to the experimental energy of 
the $1^+$-resonance, whereas in our case the experimental energies
are better reproduced by the energies for which the corresponding phase shifts are 
equal to $\pi/2$. As seen in the last column in Table~\ref{tab10Li}, when
computed in this way,
our potential gives rise to 1$^+$ and $2^+$-energies equal to 
0.37 MeV and 0.78 MeV, respectively, as well as to a $4^-$-energy of
4.88 MeV.

The differences between resonance energies obtained through the
different mathematical definition reflect an intrinsic uncertainty,
which only can be resolved by comparing calculations of directly
measured observables like specific scattering cross sections.  In this
connection, it is important that the $^9$Li-neutron interaction used
in the present work also leads to reproduction of the experimental
excitation energy spectrum of $^{10}$Li after the breakup reaction,
$d(^9\mbox{Li},p)^{10}$Li, initiated by a $^{9}$Li laboratory energy
of $11.1$~MeV/A, see Ref.\cite{mor19}.

\subsection{$^{10}$N-properties}

\begin{table}
\begin{scriptsize}
\begin{center}
\begin{tabular}{|c|cc|cc|cc|cc|c|}
\hline
  &\multicolumn{2}{c|}{This work} &\multicolumn{2}{c|}{Ref.\cite{aoy97}} &\multicolumn{2}{c|}{Ref.\cite{lep02}$^*$} &\multicolumn{2}{c|}{Ref.\cite{hoo17}}  & $\delta(E_R)=\frac{\pi}{2}$ \\ \hline
       &  $E_R$ &  $\Gamma_R$  &  $E_R$  &  $\Gamma_R$  &  $E_R$  &  $\Gamma_R$ & $E_R$ & $\Gamma_R$ & $E_R$ \\ \hline
       1$^-$ &   --    &  --    &   --  & --   &  --  & --&  $1.9^{+0.2}_{-0.2}$ & $2.5^{+2.0}_{-1.5}$ & --  \\
       2$^-$ &  1.74 &  3.94             &  1.51   &  3.47  &  --  &  -- &  $2.2^{+0.2}_{-0.2}$ & $3.1^{+0.9}_{-0.7}$ & -- \\
       1$^+$ &  2.62 & 1.68             &  2.84  &  1.89  & $2.6^{+0.4}_{-0.4}$ & $2.3^{+1.6}_{-1.6}$ &  --  &  -- & 3.51 \\
       2$^+$ & 2.89&  2.21              & 3.36   &  2.82 &   --   &  --  &   --  &  -- & 4.45 \\ \hline
\end{tabular}
\end{center}
\end{scriptsize}
\caption{For $^{10}$N, energies and widths, in MeV, of the 1$^-$, 2$^-$, $1^+$, and $2^+$ resonances obtained in our calculation (second column),  the theoretical values 
given in Ref.\cite{aoy97} (third column), and the experimental values given in Refs.\cite{lep02,hoo17} (fourth and fifth columns). The last column gives the resonance energies computed as $\delta(E_R)=\pi/2$. (*) Although in \cite{lep02} the observed resonance was assigned to be an $s$-wave resonance, as indicated
in Ref.\cite{til04}, it is very likely the energy and width corresponding to the $1^+$ state. }
\label{tab1}
\end{table}

The mirror nucleus, $^{10}$N, is now assumed to have exactly the same
potentials as $^{10}$Li, except for the Coulomb interaction arising from the
substituted valence-proton. We assume point-like protons and a
spherical and homogeneously charged $^9$C-core.  With these
interactions we now compute the spectrum of $^{10}$N as described by a
$^9$C-core and a proton.  The immediate consequence of the Coulomb
repulsion is that all the core-nucleon potentials are pushed up in
energy, reducing the depth of the potentials, and increasing the
potential barriers. This is precisely as seen by comparing the dashed and
solid curves in Fig.~\ref{figpot}.

For the $s_{1/2}$-states the Coulomb barrier implies that $1^-$ and
$2^-$-states in principle might appear as resonances.  However, the
overall repulsive behaviour of the $1^-$-potential (dashed red curve
in Fig.~\ref{figpot}) does not exhibit any barrier, and only $2^-$
resonant-states are then possible.  As for $^{10}$Li, resonances in
$^{10}$N are obtained after a complex scaling calculation.  The
results are shown in Fig.~\ref{fig1}b, where the resonances are the
isolated points out of the cut (line) associated with the continuum states.
The ground state, the $s$-wave $2^-$-state, is clearly broader than
the 1$^+$ and $2^+$ $p$-states, and therefore requires a larger angle
in the complex scaling transformation in order to be captured in the
calculation.  In particular, the calculation shown in the figure has
been made using a complex scaling angle equal to 0.5 rads for the
$2^-$-state, and 0.3 rads for the $1^+$ and $2^+$-states.

The resonance energies and widths obtained for $^{10}$N are collected
in the second column of Table~\ref{tab1}, where the results are
compared to the values given in Refs.\cite{aoy97,lep02,hoo17}.  Our
results are very consistent with the theoretical values given in
\cite{aoy97}, where the complex scaling method also is used.  The
slightly different energies are due to a core-nucleon strong
interaction producing also slightly different energies for the
$^{10}$Li-states, see Table~\ref{tab10Li}.  The experimental value in
Ref.\cite{lep02} was initially assigned to a 1$^-$-state, but in
\cite{til04} it is suggested that this resonance is very likely the
mirror of the probable $1^+$-state at 0.24~MeV in $^{10}$Li.

Finally, in Ref.\cite{hoo17} (fifth column in Table~\ref{tab1}) two
resonances have been measured with energies around 2.0~MeV, which are
assigned by the authors to states with angular momentum and parity, $1^-$ and
$2^-$.  In this reference the authors mention as well an excited $1^-$
or $2^-$ resonant-state with an energy of $2.8\pm 0.2$~MeV.  As seen
in Fig.~\ref{fig1}b, we have not found any trace of such an excited
state with negative parity.  It is in any case striking, that the three
energies reported in \cite{hoo17} agree reasonably well with the three
energies obtained in this work for the $2^-$, $1^+$, and $2^ +$-states.

Due to its large resonance width, the $2^-$-phase shift never reaches
the value of $\pi/2$, and therefore the energy of this resonance can
not be extracted in this way.  In contrast, this is possible for the
$1^+$ and $2^+$-states, and this happens for energies equal to 3.51
MeV and 4.45 MeV, respectively, that is clearly larger than the
energies obtained as poles of the $S$-matrix (last column in
Table~\ref{tab1}).  Again these different definitions reflect in
themselves an inherent uncertainty in the resonance parameters.

\section{The core-nucleon-nucleon system}

After discussing the two-body properties of $^{10}$Li and $^{10}$N, we
now investigate the effects they determine for the structure and
properties of the three-body mirror nuclei, $^{11}$Li and $^{11}$O.
We shall do this in three different steps, first in this section we
present the energy spectra, and in the two following sections we
discuss the properties of the different states, respectively the
internal structure of the wave functions, and the spatial distribution
of the three constituents.

\subsection{Energy spectrum of $^{11}$Li}

\begin{figure}
\begin{center}
\includegraphics[width=\linewidth,angle=0]{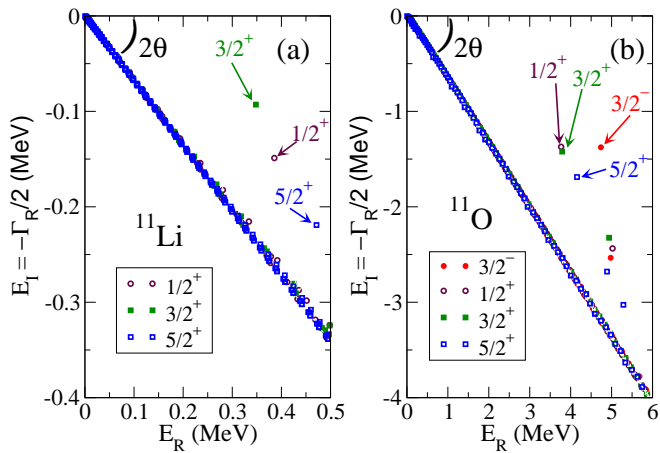}
\end{center}
\caption{Complex resonance energies for the computed $1/2^+$ (open brown circles), $3/2^+$ (solid green squares), and $5/2^+$ (open blue squares) states in 
$^{11}$Li, panel (a), and $^{11}$O, panel (b), where the $3/2^-$resonance (solid red circles) is also shown. The calculations have been performed with a complex scaling
angle $\theta=0.3$ rads.}
\label{fig2}
\end{figure}

The potentials used to describe $^{11}$Li are the same as
used in Ref.\cite{cas17}, where the properties of the computed $3/2^-$
ground-state wave function in $^{11}$Li are described. An effective
three-body force is used to fit the experimental two-neutron
separation energy of $369.15(65)$~keV \cite{smi08}, which leads to a
charge root-mean-square radius of 3.42 fm, also in agreement with the
experimental value reported in \cite{smi08}. In this work an attractive Gaussian 
three-body force with a range of 5 fm and a strength of $-0.6$ MeV has 
been used.  In addition, as shown in
Ref.\cite{cas17}, the computed $^{11}$Li ground-state wave function
permits reproduction of the experimental energy-integrated angular
differential cross section for the $^{11}$Li$(p, d)^{10}$Li reaction
at 5.7 MeV/A.

In Ref.\cite{gar02} the electric dipole excitations in $^{11}$Li,
i.e. the $1/2^+$, $3/2^+$, and $5/2^+$ resonant-states, were
investigated by means of the complex scaling method
\cite{ho83,moi98}. It was found that the energies of these three
resonances are pretty close to each other, with specific values
depending slightly on the properties of the core-neutron interaction.
In any case, they lie in the energy interval between 0.3~MeV and
0.7~MeV above the three-body threshold.

\begin{table}
  \begin{tabular}{|c|c|c|c|}
\hline
       &  $^{11}$Li (Comp.) &  $\Delta_c^{(1)}$ & $^{11}$O (Estim.)\\ \hline
$\frac{3}{2}^-$ & $-0.37$            &       4.93           &    4.56 \\
$\frac{1}{2}^+$ & $0.39-i0.15$   &   $2.69-i0.77$ &  $3.08-i0.92$  \\
$\frac{3}{2}^+$ & $0.35-i0.09$   &   $3.41-i0.86$&   $3.76-i0.95$\\
$\frac{5}{2}^+$ & $0.47-i0.22$   &   $3.89-i1.20$&  $4.36-i1.42$  \\ \hline
\end{tabular}
\caption{Computed complex energies, $E_R-i\Gamma_R/2$, of the $\frac{3}{2}^-$,  $\frac{1}{2}^+$, $\frac{3}{2}^+$, and $\frac{5}{2}^+$ states in $^{11}$Li, the Coulomb shift for each of them as defined in Eq.(\ref{eq2}), and the estimated complex energies of the corresponding states in $^{11}$O. All the values are given in MeV.}
\label{tab0}
\end{table}

When the specific interactions used in this work are employed, the
complex scaling method reveals the existence of $1/2^+$, $3/2^+$, and
$5/2^+$-resonances.  The results are shown in Fig.~\ref{fig2}a, where
specific dots appear clearly separated from the straight line
corresponding to the background continuum states.  The resonances of
interest are indicated by the arrows in the figure.  The precise
computed values for the resonance energies and widths, $(E_R,
\Gamma_R)$, are $(0.39, 0.30)$ MeV, $(0.35, 0.18)$ MeV, and $(0.47,
0.44)$ MeV for the $1/2^+$, $3/2^+$, and $5/2^+$-states, respectively.
This is also given in the second column in Table~\ref{tab0} as a
complex number for each state.

\subsection{Energy spectrum of $^{11}$O}

As expected, due to the Coulomb repulsion, the ground state in $^{11}$O
is not bound. Therefore, in this case all the states, $J^\pi=3/2^ -$,
$1/2^ +$, $3/2^ +$, and $5/2^+$, will be computed by means of the
complex scaling method. We have used a complex scaling angle of
$\theta=0.30$ rads, and the result of the calculation is shown in
Fig.~\ref{fig2}b. The straight line, rotated by an angle equal to
$2\theta$, contains the discretized continuum states, and the points
out of this line correspond to the different resonances. The lowest
$3/2^-$, $1/2^+$, $3/2^+$, and $5/2^+$-states are indicated by the
corresponding arrows.

As seen in the figure, in all the cases a second resonance is found in
the vicinity of $E_R=5$ MeV. In the 5/2$^+$-case even a third
resonance around 5.5 MeV is seen. In order to make the plot clean, the
cuts associated to the two-body resonances, i.e. two-body resonance
plus the third particle in the continuum \cite{moi98}, are not shown
in the figure.

\begin{table}
  \begin{tabular}{|c|cc|cc|cc|cc|}
    \hline
   & \multicolumn{2}{c|}{This work} & \multicolumn{2}{c|}{Ref.\cite{web19,wan19}} & \multicolumn{2}{c|}{Ref.\cite{cha12}} & \multicolumn{2}{c|}{Ref.\cite{for19}}  \\ \hline
   &  $E_R$  &  $\Gamma_R$   &  $E_R$  &  $\Gamma_R$&  $E_R$  &  $\Gamma_R$ &  $E_R$  &  $\Gamma_R$  \\ \hline
 $\frac{3}{2}^-$   & 4.74 & 2.75  & 4.16  & 1.30  & $3.21\pm 0.84$  & -- & 4.75   & 2.51      \\
                            &  4.97  & 5.07 & 4.85 &  1.33  &     --                                  &--  & --       &--            \\ \hline
 $\frac{1}{2}^+$ & 3.77  &2.74 &          &           &     --                                  &--  & --       &--           \\ 
                           &  5.02  &4.87  &         &         &     --                                  &--  & --       &--           \\ \hline    
 $\frac{3}{2}^+$ & 3.79  & 2.84&         &           &     --                                  &--  & --       &--           \\ 
                           &  4.93  & 4.65&         &         &     --                                  &--  & --       &--           \\ \hline                                                    
 $\frac{5}{2}^+$ & 4.16  & 3.38& 4.65 &  1.06  &     --                                  &--  & --       &--           \\ 
                           &  4.89  &  5.36 & 6.28  & 1.96  &     --                                  &--  & --       &--           \\ \hline
 \end{tabular}
\caption{For $^{11}$O, the second column gives the computed energies, $E_R$, and widths, $\Gamma_R$, of the $3/2^-$, $1/2^+$, $3/2^+$ and $5/2^+$states. The last 
three columns show the results given in Refs.\cite{web19,wan19}, Ref.\cite{cha12}, and Ref.\cite{for19}, respectively. Both, energies and widths, are given in MeV.}
\label{tab3}
\end{table}

The precise values of the resonant energies, $E_R$, and widths,
$\Gamma_R$, are given in the second column of Table~\ref{tab3} for the
two lowest resonances for each of the computed $J^ \pi$-states.   
As seen in the table, the $1/2^+$, $3/2^+$,
and $5/2^+$-energies are similar to each other, especially the 1/2$^+$
and $3/2^+$-states, which are almost degenerate. Therefore, very
likely one of these states should actually be the ground state, since
the $3/2^-$-energy is at least 0.6~MeV higher.

At this point, we emphasize that the computed resonance energies and widths  have
been obtained without inclusion of any three-body force in the radial
Eqs.(\ref{radf}).  When a short-range effective three-body potential
is included, we have observed that the effect is clearly bigger for
the 3/2$^-$-state than for the positive parity resonances.  This is an
indication that for the $3/2^-$-case the core and the two
valence-protons are clearly closer to each other than for the $1/2^+$,
$3/2^+$, and $5/2^+$-states.

In particular, if we use a Gaussian three-body force with a range of 5
fm and an attractive strength of $-4$~MeV, the energies and widths of
the positive parity resonances remain essentially unchanged, whereas
for the lowest $3/2^-$-state we get an energy of 3.72 MeV (with a width of
1.18 MeV), similar to the one of the other resonances.  The conclusion
is then, that from the pure three-body calculation it is difficult, or
even not possible, to determine which $J^\pi$-state is actually the
ground state in $^{11}$O. In any case, it seems clear that we cannot
exclude the possibility that another state than the $3/2^-$-state
becomes the ground state in $^{11}$O.

In Table~\ref{tab3} we compare our results with previous works. In all
of them, the ground state is assigned to an angular momentum and
parity, $3/2^-$. Only in Refs.\cite{web19,wan19} the positive parity
state $5/2^+$ is also considered. In these references,
\cite{web19,wan19}, the computed resonances are clearly narrower than
the ones obtained in the present work, and the fact that the two
energies given for the lowest 3/2$^-$ and $5/2^+$-states are similar
to ours, but with the quantum numbers exchanged, is probably just an
accidental coincidence. 
The reason for this difference is difficult to determine.
On the one hand, in our work the core is assumed to be spherical in
contrast to Refs.\cite{web19,wan19}. Thus, we have ignored the
possibly significant role played by the deformation.  On the other
hand, in Refs.\cite{web19,wan19} the calculations are performed taking
a maximum value for the hypermomentum, $K_{max}$, equal to $20$.  As
we shall discuss later, this value might be too small to guarantee
that convergence has been reached in the calculations.  

Our three-body
approach treats the constituents as inert particles with central
two-body interactions, and it is therefore insensitive to deformation.
In any case, 
as discussed in \cite{mis97}, for sufficiently weakly bound systems, or
in other words, provided the valence nucleons are located at relatively large 
distance from the core,  ``the quadrupole deformation of the resulting halo 
is completely determined by the intrinsic structure of a weakly bound 
orbital, irrespective of the shape of the core''.
Furthermore, our phenomenological choice of interaction
parameters necessarily accounts for at least part of the effects of the
core-deformation.

A further comparison is found in another publication \cite{cha12},
where a clearly lower energy, 3.21 MeV, although with a large error bar, is given for the
$3/2^-$-state.  However, this energy has not been computed, but
obtained as an extrapolation using the isobaric multiplet
mass-equation, whose coefficients are determined after the shell-model
computed energies for $^{11}$Li, $^{11}$Be, and $^{11}$B.  In addition
to these energy properties, it was argued in Ref.\cite{for19}, that
the experimental breakup data given in \cite{web19}, can be as well
reproduced by only the ground state of $^{11}$O, whose energy and
width are given in Table~\ref{tab3} to be 4.75 MeV and 2.51 MeV,
respectively, very similar to our lowest $3/2^-$ energy and width.

\subsection{Coulomb shift}

Due to the charge symmetry of the strong interaction, the
$^{11}$O-states computed in the previous subsection have been obtained
simply by adding the Coulomb potential to the nuclear interactions
used to describe $^{11}$Li. The repulsive character of the Coulomb
interaction, between the two valence-protons and between each of the
valence-protons and the core, is obviously the reason for the increase
of the energies.

An estimate of how much these energies should be modified by the Coulomb repulsion can be obtained as the first-order perturbative value of the Coulomb shift:
\begin{equation}
\Delta_c^{(1)}=\langle \Psi(\mbox{$^{11}$Li}) | V_{coul}| \Psi(\mbox{$^{11}$Li}) \rangle,
\label{eq2}
\end{equation} 
where $ \Psi(\mbox{$^{11}$Li})$ is the three-body $^{11}$Li wave
function corresponding to a given state, but where the
valence-neutrons are replaced by protons, and the charge of the core
is assumed to be the one of the mirror nucleus, $^9$C.  Therefore,
$\Psi(\mbox{$^{11}$Li})$ in Eq.(\ref{eq2}) represents an artificial
$^{11}$O wave function, which is assumed to have the same structure as
the corresponding $^{11}$Li-state. The $V_{coul}$-potential is the
resulting Coulomb interaction between the three pairs of charged
particles, where the protons are point-like and the $^9$C-core is
spherical and uniformly charged corresponding to the root-mean-square
radius of about $2.5$ fm.  The first order Coulomb shift,
$\Delta_c^{(1)}$, is then the diagonal contribution to the Coulomb
shift.

As already mentioned,  the $1/2^+$, $3/2^+$, and $5/2^+$-states in $^{11}$Li have been obtained by means of the complex scaling method. The wave functions are then complex
rotated, and the corresponding value of $\Delta_c^{(1)}$ has to be obtained after complex rotation of the Coulomb potential. In this way $\Delta_c^{(1)}$ will be a complex quantity, whose imaginary part can be interpreted as the uncertainty in the energy shift \cite{moi98}. In other words, the imaginary part of $\Delta_c^{(1)}$ in the $1/2^+$, $3/2^+$, and $5/2^+$-cases permits us to estimate as well the change in the width of the resonance.

In Table~\ref{tab0} we give the computed energies of the
$^{11}$Li-states (second column) together with the computed values of
$\Delta_c^{(1)}$ for the four states considered (third
column). The results depend slightly on how the Coulomb potential is
constructed, but the overall size and relations are very well
determined.  As we can see, the value of $\Delta_c^{(1)}$ is
substantially larger for the $3/2^-$-state than for the other three
states, which again is an indication of the smaller size of the
$3/2^-$-state, since the smaller the system the larger the Coulomb
repulsion, and therefore the larger the value of $\Delta_c^{(1)}$.

When this shift is added to the $^{11}$Li-energies, we obtain the
estimate for the energies of the $^{11}$O-states given in the last
column of the table. In the case of the $3/2^-$-state, since the
$^{11}$Li wave function is real, the shift $\Delta_c^{(1)}$ is also
real, and an estimate of the width in the $3/2^-$-state in $^{11}$O is
not possible in this way.  As we can see, the estimated energies given
in the last column of Table~\ref{tab0} are quite reasonable, pretty
close to the computed energies given in Table~\ref{tab3} for the
lowest $3/2^-$, $1/2^+$, $3/2^+$, and $5/2^+$ states.  The only
exception is perhaps the $1/2^+$-state, where a difference of about
0.7 MeV is found.  

These similarities show that the
variations in the energy shift due to the structure
differences, as expected, are relatively small.
The difference between the energy shift, $\Delta_c^{(1)}$, for a given
state and the experimental energy shift, $\Delta_c$, between the
experimental energies is a measure of the structure effect of the
Thomas-Ehrman shift, that is $\Delta_{TE}=\Delta_c - \Delta_c^{(1)}$.
Recent calculations of $\Delta_{TE}$ concerning different light
mirror nuclei are available in the literature. For example, in
Ref.\cite{aue00} the shift between the mirror system, $^{11}$Be and
$^{11}$N, was investigated. In \cite{gri02} the same was done for
$^{12}$Be-$^{12}$O and $^{16}$C-$^{16}$Ne, and in Ref.\cite{gar04}
this shift was computed for the case of $^{17}$N and $^{17}$Ne. In all
the cases the value of $\Delta_{TE}$ was obtained to be of no more
than a few hundreds of keV. These sizes are consistent with the
energy difference between the lowest $J^\pi$-energies obtained in our
calculation (Table~\ref{tab3}), and the estimated energies given in
the last column of Table~\ref{tab0}.

\section{Three-body wave functions}

\begin{table}
  \begin{tabular}{|c|cccccc|cc|}
    \hline
  &  \multicolumn{6}{c|}{Component}  &\multicolumn{2}{c|}{\%} \\ \hline
  $J^\pi$ & $\ell_x$ & $\ell_y$ & $L$ & $s_x$ &  $S$  &  $K_{max}$  & $^{11}$Li & $^{11}$O\\ \hline
   $\frac{3}{2}^-$  &  0  &  0   &   0    &    1 &   $3/2$  &  120   &22\% & 5\% \\
                             &   0  &  0   &   0    &   2  &   $3/2$  &  120   &35\% & 7\% \\
                             &   1  &  1   &   0    &   1  &   $3/2$  &  60     & 6\% & 11\% \\
                             &   1  &  1   &   0    &   2  &   $3/2$  &  80     &10\% &19\% \\
                             &   1  &  1   &   1    &   1  &   $1/2$  &  60     & 4\% & 9\% \\
                             &   1  &  1   &   1    &   1  &   $3/2$  &  40     &5\% & 11\% \\
                             &   1  &  1   &   1    &   2  &   $3/2$  &  40    &3\% & 7\% \\
                             &   1  &  1   &   1    &   2  &   $5/2$  &  80    &12\% &28\% \\ \hline
  $\frac{1}{2}^+$ &   0  &  1   &   1    &   1  &   $1/2$  &  121  & $<1$\% &  5\%\\
                             &   0  &  1   &   1    &   1  &   $3/2$  &  201  &30\% & 37\% \\
                             &   1  &  0   &   1    &   1  &   $1/2$  &  121  &6\% & 6\% \\
                             &   1  &  0   &   1    &   1  &   $3/2$  &  121  & 7\% & 6\% \\                            
                             &   1  &  0   &   1    &   2  &   $3/2$  &  201  &42\% & 41\% \\
                             &   1  &  2   &   1    &   1  &   $3/2$  &    61  & 7\% & 5\% \\
                             &   2  &  1   &   1    &   1  &   $3/2$  &    41  & 1\% & $<1$\% \\ \hline
 $\frac{3}{2}^+$  &   0  &  1   &   1    &   1   &  $1/2$  &  101  &1\% & $<1$\% \\
                             &   0  &  1   &   1    &   1   &  $3/2$  &  101  &1\% & 1\%\\
                             &   0  &  1   &   1    &   2  &   $3/2$  &  161  &9\% & 7\% \\
                             &   0  &  1   &   1    &   2  &   $5/2$  &  201  &34\% &  32\% \\
                             &   1  &  0   &   1    &   1  &   $1/2$  &  101  &1\% & 1\% \\
                             &   1  &  0   &   1    &   1  &   $3/2$  &  161  &8\% & 7\% \\
                             &   1  &  0   &   1    &   2  &   $3/2$  &  161  &5\% &4\%\\
                             &   1  &  0   &   1    &   2  &   $5/2$  &  201  &36\% &43\% \\ 
                             &   1  &  2   &   1    &   2  &   $5/2$  &   41   & 2\% & 4\% \\   \hline     
$\frac{5}{2}^+$   &   0  &  1   &   1    &   2  &   $3/2$  &  201  &24\% & 23\%\\
                             &   0  &  1   &   1    &   2  &   $5/2$  &  201  &25\% & 20\% \\
                             &   1  &  0   &   1    &   1  &   $3/2$  &  201 &25\% & 26\% \\
                             &   1  &  0   &   1    &   2  &   $3/2$  &   81  &3\% & 2\% \\
                             &   1  &  0   &   1    &   2  &   $5/2$  &  201  &23\% & 25\% \\
                             &   1  &  2   &   1    &   2  &   $3/2$  &   31  & $<1$\% & 2\% \\
                             &   1  &  2   &   1    &   2  &   $5/2$  &   31  & $<1$\% &  2\% \\ \hline                                                 
\end{tabular}
\caption{Dominant components (larger than 1\% probability) in the
  lowest $3/2^-$, $1/2^+$, $3/2^+$, and $5/2^+$ wave functions in
  $^{11}$Li and $^{11}$O in the Jacobi set with the
  $\bm{x}$-coordinate defined between the core and one of the
  valence nucleons. The quantum numbers are as defined below Eq.(\ref{exp1}).
  Note that the core has negative parity.}
\label{tabcom0}
\end{table}

The calculation of the $^{11}$Li and $^{11}$O three-body states has
been made including all the components satisfying $\ell_x,\ell_y \leq
7$, where $\ell_x$ and $\ell_y$ are the relative angular momenta
between two of the particles, and between their center-of-mass and the
third particle, respectively.  The maximum value of the hypermomentum,
$K_{max}$, has to be sufficiently large to reach convergence, but for
all partial waves it has been taken to be at least $20$.  In Table~\ref{tabcom0} we
give the partial wave decomposition and the components with probability
larger than 1\% for each of the lowest $J^\pi$-states.  Note here that the
use of the complex scaling method permits us to normalize the resonance
wave functions as described in Ref.\cite{moi98}.  The quantum numbers
in the table are as described below Eq.(\ref{exp1}), with the
$\bm{x}$-Jacobi coordinate defined between the core and one of the
valence nucleons.

As seen in the table, for these components the $K_{max}$-value used is
pretty large, very likely overdoing the work of getting a
well-converged three-body solution, especially for $^{11}$Li. A careful
analysis of each individual component could certainly reduce the
$K_{max}$-value. As a test, we have performed the same calculations
using $K_{max}=20$ for all the components. These less accurate
calculations result in an increase of the three-body energies by up to
0.4 MeV for the $^{11}$Li-states and by up to 1 MeV for the $^{11}$O-resonances. The only exception is the lowest $3/2^-$-state (bound in
the case of $^{11}$Li), for which the increase in energy is of about
50 keV for $^{11}$Li, and of about 200~keV for $^{11}$O. This is once
more reflecting the smaller size of the lowest $3/2^-$-state compared
to the positive parity states, since the closer the particles are to each
other, the smaller is the basis necessary to reach convergence.

\begin{table}
  \begin{tabular}{|c|ccccc|cc|}
    \hline
    &  \multicolumn{5}{c|}{Component} & \multicolumn{2}{c|}{\%} \\ \hline
  $J^\pi$ & $\ell_x$ & $j_N$ & $j_x$ & $\ell_y$ &  $j_y$   &  $^{11}$Li  & $^{11}$O \\ \hline
   $\frac{3}{2}^-$  &  0  &  1/2   &   1    &   0  &   $1/2$  & 22\% & 5\%\\
                             &   0  &  1/2   &   2    &   0  &   $1/2$  & 35\% & 7\%\\
                             &   1  &  1/2   &   1    &   1  &   $1/2$  & 15\% & 32\%\\
                             &   1  &  1/2   &   2    &   1  &   $1/2$  & 25\% & 53\% \\ \hline
  $\frac{1}{2}^+$ &   0  &  1/2   &   1    &   1  &   $1/2$  & 34\% & 41\%\\
                             &   1  &  1/2   &   1    &   0  &   $1/2$  & 52\% & 51\%\\
                             &   1  &  1/2   &   2    &   2  &   $3/2$  & 3\% & 2\% \\
                             &   1  &  3/2   &   0    &   0  &   $1/2$  & 1\% & 1\% \\ 
                             &   1  &  3/2   &   1    &   0  &   $1/2$  & 2\% & 1\% \\
                             &   1  &  3/2   &   1    &   2  &   $3/2$  & 1\% & 1\% \\
                             &   1  &  3/2   &   2    &   2  &   $3/2$  & 3\% & 2\% \\ \hline
 $\frac{3}{2}^+$  &   0  &  1/2   &   1    &   1  &   $1/2$  & 2\% &  2\% \\
                             &   0  &  1/2   &   2    &   1  &   $1/2$  & 45\% & 39\% \\
                             &   1  &  1/2   &   1    &   0  &   $1/2$  & 51\% & 52\% \\ 
                             &   1  &  1/2   &   3    &   2  &   $3/2$  & $<1$\%  &  3\%\\ \hline    
  $\frac{5}{2}^+$ &   0  &  1/2   &   2    &   1  &   $1/2$  & 49\% & 44\% \\
                             &   1  &  1/2   &   1    &   2  &   $3/2$  &  $<1$\% &  1\%\\
                             &   1  &  1/2   &   2    &   0  &   $1/2$  & 49\% & 52\% \\
                             &   1  &  1/2   &   2    &   2  &   $3/2$  &  $<1$\% &  1\% \\ \hline                                     
\end{tabular}
\caption{The same as Table~\ref{tabcom0} but in the coupling scheme where the core-neutron relative orbital angular momentum $\ell_x$ couples to the spin of the neutron
to provide the angular momentum $j_N$, which couples to the spin of the core to the total core-neutron angular momentum $j_x$. The relative orbital angular momentum
between the core-neutron center-of-mass and the second neutron, $\ell_y$, couples to the spin of the second neutron to give the angular momentum $j_y$. Both, $j_x$
and $j_y$ couple to the total three-body angular momentum $J$.}
\label{tabcom2}
\end{table}

The same decomposition is shown in Table~\ref{tabcom2}, but in a
coupling scheme more consistent with the mean-field quantum numbers,
where the core-nucleon relative orbital angular momentum, $\ell_x$,
couples to the spin of the nucleon to provide the angular momentum
$j_N$, which in turn couples to the spin of the core to give the total
core-nucleon angular momentum, $j_x$. The relative orbital angular
momentum between the core-nucleon center-of-mass and the second
nucleon, $\ell_y$, couples to the spin of the second nucleon to give
the angular momentum, $j_y$. Both $j_x$ and $j_y$ couple to the total
three-body angular momentum, $J$.

As we can see, the structure of the $3/2^-$-state changes
substantially due to the Coulomb repulsion. In the case of $^{11}$Li
the $3/2^-$ (bound) ground state contains about 40\% of core-neutron
$p$-wave contribution. More precisely, 15\% of the wave function
corresponds to $^{10}$Li in the $1^+$-state, and 25\% to $^{10}$Li in
the $2^+$-state (Table~\ref{tabcom2}). With respect to the $s$-wave
contribution, even if the $1^-$-state in $^{10}$Li is lying high in the
continuum, 22\% corresponds to $^{10}$Li populating that state.

These characteristics are however very different when analyzing the
$3/2^-$-state in $^{11}$O.  The Coulomb repulsion, which pushes up the
$s$-wave core-proton 2$^-$-state by more than 1.5 MeV, turns out to be
crucial producing a drastic reduction of the $s$-wave
contribution. In fact, as seen in the upper part of
Tables~\ref{tabcom0} and \ref{tabcom2}, the $p$-wave components give
85\% of the wave function, whereas the $s$-wave contribution reduces
now from almost 60\% in $^{11}$Li to only about 12\% in $^{11}$O.

This result is in contrast to Ref.\cite{wan19}, where the $29\%$
$s$-wave contribution in the $3/2^-$-state in $^{11}$O is even higher
than the 25\% given for $^{11}$Li.  This low $s$-wave content in the
$^{11}$Li ground-state wave function seems to disagree with previous
results in \cite{gar96,gar97,cas17}, where it is shown that the
agreement with experimental momentum distributions and angular
differential cross sections requires a $p$-wave content of about
$35\%-40\%$ in the $^{11}$Li ground-state, or, equivalently,
$60\%-65\%$ $s$-wave contribution.

For both nuclei, $^{11}$Li and $^{11}$O, the contribution of $d$-waves
in the present work is far from substantial, in total of about 3\% in
both cases, and none of the $d$-wave components provides more than 1\%
of the norm.  This result seems to contradict the measured increase of
about 8.8\% \cite{neu08} of the quadrupole moment in $^{11}$Li
relative to that of $^{9}$Li, which in shell-model calculations in
Ref.\cite{suz03} is explained as due to a significant $d$-wave
contribution of similar size as the one corresponding to $p$-waves.

The small probability of $d$-waves may be related to the lack of deformation of the frozen core as seen by the argument.  If the structure of a given deformation is expanded on another, say body-fixed, deformation, there must be partial wave components corresponding to this deformation.  However, our three-body model provides the full wave function corresponding to that obtained with deformation after, not before, projection of angular momentum and parity. Thus, our model can only say something about the inserted frozen core-structure and the calculated valence-structure.

However, as discussed in \cite{neu08}, the measured quadrupole
moment could instead be related to an about $10\%$ increase of the
charge-radius in $^{11}$Li compared to the one of $^{9}$Li.  This
increase can be interpreted as due to the neutron halo.
The two neutrons 
in the zero angular momentum ground state produce a
distortion of the $^{9}$Li-core, which effectively corresponds to an
increase of the core-radius.  The initially slightly deformed
$^{9}$Li-core is otherwise in principle maintained in the subsequent
three-body calculations without significant effect as argued in
\cite{mis97}.  Using such an increased radius, the neutron-core
interaction still must be adjusted to the previously described
specific desired properties.  These all-decisive phenomenologically
obtained interactions guarantee the same resulting three-body
structure as obtained with the bare $^{9}$Li-radius.  This is
consistent with \cite{neu08}, where it is stated that there is a
striking analogy between the quadrupole moment and the
root-mean-square charge-radius without any additional change of wave
function structure.

In any case, a detailed analysis of the quadrupole moment of $^{11}$Li 
requires to take into account the different sources contributing to
the measured value, since the two valence neutrons are mostly on the same side 
of $^{9}$Li.  First the contribution from the original $^{9}$Li  quadrupole moment, 
second the one from the rotation of the $^9$Li core around
the three-body center of mass,  and third and fourth the one from increased radius and induced 
deformation of  $^{9}$Li from the two valence neutrons.  From Ref.\cite{boh75}  we 
know that  neutral nucleons polarize the charged core by an amount 
of the same order as if they were charged. Thus, our model is consistent with all available 
$^{11}$Li data, but for now we leave the complicated quantitative quadrupole moment 
calculation for future investigations.

Concerning the $1/2^+$, $3/2^+$, and $5/2^+$-resonances, they are all
almost completely given by $sp$-interference terms. Only minor
contributions from $pd$-interferences are seen, with the largest, as
given in the tables, reaching up to 7\% in the $1/2^+$-case.  
The presence of low-lying $p$-resonances in $^{10}$Li and $^{10}$N makes
the $pd$-interferences  more likely than the $dp$-ones, whose weight
is always smaller than 1\%. 
In general, we can see that the structure of the three positive parity
resonances does not change significantly by moving from $^{11}$Li to
$^{11}$O.  The weight of the different components remains essentially
the same in both cases. 

An important difference, seen in Table~\ref{tabcom2}, between the
structure of the different $J^+$-resonances, is that only the
$1/2^+$-state has substantial contributions from $s$-waves
($\ell_x=0$) in the nucleon-core $1^-$-state ($j_x=1$) in either $^{10}$Li or
$^{10}$N.  In contrast, for both the $3/2^+$ and $5/2^+$-resonances
only the 2$^-$ $s$-state ($\ell_x=0$, $j_x=2$) in $^ {10}$Li or
$^{10}$N is substantially populated.  In the next section, we shall
discuss this difference as responsible for the very different spatial
structure of these states. Keep in mind that  a similar weight of the different
partial-wave components does not necessarily imply a similar
spatial distribution of the constituents, which is in fact determined by
the radial wave functions $f_n^J(\rho)$ in Eq.(\ref{exp0}) and 
the expansion coefficients $C_q^{(n)}(\rho)$ in Eq.(\ref{exp1}).
A different $\rho$-dependence can still provide a similar weight
of the partial waves after integration of the square of the wave function.

\section{Three-body spatial structure}

Let us examine now the spatial structure of the $^{11}$Li and
$^{11}$O-states. A clean indication of how the core and valence
nucleons are distributed in space is reflected by the different
two-body root-mean-square (rms) radii, which in turn permit us to
obtain a clear picture of the most probable inter-particle distances.
We therefore first discuss the radii or second radial moments, and
afterwards the origin in the structures of the wave functions.

\subsection{Radii}

Since the complex scaling method has been used in the calculations, the
corresponding resonance three-body wave functions are complex rotated.
As a consequence, for the resonant states, the rms radii have to be obtained as the 
expectation value of the square of complex rotated radial distance ($r\rightarrow
re^{i\theta}$).  Therefore, the rms radii are in this case complex
quantities, and as described in \cite{moi98}, the imaginary part is a
measure of the uncertainty of the computed value.

\begin{table}
\begin{tabular}{|c|cc|cc|}
\multicolumn{5}{c}{$^{11}$Li} \\ \hline
$J^\pi$ & $\langle r_{nn}^2\rangle^{1/2}$ &  $\langle r_{c,nn}^2\rangle^{1/2}$ &  $\langle r_{cn}^2\rangle^{1/2}$ &  $\langle r_{n,cn}^2\rangle^{1/2}$ \\ \hline
$\frac{3}{2}^-$ & 6.4 &  5.0  & 5.9  & 5.7 \\
$\frac{1}{2}^+$ &$22.3+i6.3$   &  $11.7+i3.0$  &  $16.5+i4.4$ & $17.1+i4.9$ \\
$\frac{3}{2}^+$ &  $13.9+i4.8$ & $7.4+i3.4$  & $10.4+i4.1$  &  $10.7+i4.1$ \\
$\frac{5}{2}^+$ & $9.4+i4.3$  &  $3.1+i0.9$  &  $6.8+i2.0$ &   $7.0+i2.2$ \\ \hline
\multicolumn{5}{c}{ } \\
\end{tabular}
\begin{tabular}{|c|cc|cc|}
\multicolumn{5}{c}{$^{11}$O} \\ \hline
$J^\pi$ & $\langle r_{pp}^2\rangle^{1/2}$ &  $\langle r_{c,pp}^2\rangle^{1/2}$ &  $\langle r_{cp}^2\rangle^{1/2}$ &  $\langle r_{p,cp}^2\rangle^{1/2}$ \\ \hline
$\frac{3}{2}^-$ & $5.1+i 3.9$  & $2.8+i2.1$   &  $3.9+i3.0$ & $3.9+i3.0$ \\
$\frac{1}{2}^+$ & $12.8+i5.0$  &  $6.6+i3.0$  & $9.3+i4.0$  & $9.4+i4.0$ \\
$\frac{3}{2}^+$ &  $12.2+i5.4$ & $6.5+i2.9$  & $9.0+i4.1$  &  $9.0+i4.1$ \\
$\frac{5}{2}^+$ &  $10.7+i6.2$ &  $5.2+i3.1$  &  $7.5+i4.5$ &  $7.5+i4.6$ \\
\hline
\end{tabular}

\caption{ Computed values, in fm, of $\langle r_{ij}^2\rangle^ {1/2}$ and $\langle r_{k,ij}^ 2\rangle^ {1/2}$, for the $3/2^-$, $1/2^+$, $3/2^+$, and $5/2^+$ states
in $^{11}$Li (upper part) and $^{11}$O (lower part), where $\{ i, j, k\}$ represent the core ($c$) and the valence neutrons ($n$), or the core ($c$)
and the valence protons ($p$), respectively. The coordinate $r_{ij}$ is the distance between particles $i$ and $j$, and
$r_{k,jk}$ is the distance between particle $k$ and the center-of-mass of the $ij$-system.}
\label{tabr2}
\end{table}

In Table~\ref{tabr2} we give the rms radii, $\langle
r_{ij}^2\rangle^{1/2}$ and $\langle r_{k,ij}^2\rangle^{1/2}$, for the
different $^{11}$Li (upper part) and $^{11}$O (lower part) states.
From the right part of the table we notice that for all the states in
both $^{11}$Li and $^{11}$O, the distances, $\langle
r_{cN}^2\rangle^{1/2}$ and $\langle r_{N,cN}^2\rangle^{1/2}$, are
similar to each other, where $N$ can be either neutrons ($n$) or
protons ($p$).  Since the core is about nine times heavier than the
nucleon, the value of $\langle r_{N,cN}^2\rangle^{1/2}$ is not far
from the distance between the core and the second nucleon, which
implies that the two valence-nucleons are roughly at the same distance
from the core in all the $J^\pi$-states.

Looking now into the left part of Table~\ref{tabr2}, we can see that
for the bound $3/2^-$-state in $^{11}$Li the neutron-neutron distance
is similar to the core-neutron distance, which indicates an
equilateral triangular structure with a particle-particle distance of
about 6 fm.  However, for the resonant states in $^{11}$Li
and $^{11}$O the situation is slightly different, since the nucleon-nucleon
distance is roughly 1.4 times larger than the core-nucleon distance. This
structure corresponds to an isosceles triangle, where the unequal side
(the nucleon-nucleon distance) is about 40\% bigger than the two
equal sides given by the core-nucleon distance.

It is also interesting to note that the $3/2^-$-state for both nuclei
is clearly smaller than the positive-parity resonances.  This was
already anticipated by the larger effect of the three-body force,
the larger value of $\Delta_c^{(1)}$ (Table~\ref{tab0}), and the smaller
$K_{max}$-values required to get convergence for the
$3/2^-$-states.  This is related to the facts, that in $^{11}$Li the $3/2^-$-state
is bound, and in $^{11}$O the $3/2^-$ wave function has a
clearly dominant contribution from two valence-protons in a relative
$p$-wave with respect to the core (85\% according to
Table~\ref{tabcom2}). In this structure the potential barrier prevents
the protons from moving too far away from the core (see the dashed
curves in Fig.~\ref{figpot}b).

On the other hand, as mentioned when discussing Table~\ref{tabcom2},
the $1/2^+$, $3/2^+$, and $5/2^+$-resonances are almost entirely
$sp$-structures, which implies that one of the halo nucleons is always
populating a core-nucleon $s$-state.  As seen in Fig.~\ref{figpot}a,
the $s$-wave potentials do not feel any confining barrier, except the
$2^-$-potential in $^{10}$N (dashed black curve), for which the
potential barrier is almost a factor of 2 lower than for the
$p$-potentials for the same system. As a consequence, the 
positive-parity resonances are, as seen in Table~\ref{tabr2},
significantly bigger than for the 3/2$^-$-states.

Also, as already mentioned and shown in Table~\ref{tabcom2}, 
the contribution of the nucleon-core $1^-$-state ($\ell_x=0$, $j_x=1$) 
to the $J^+$-resonances is only substantial for the $1/2^+$-state, 
whereas for the $3/2^+$ and
$5/2^+$-resonances basically all the $\ell_x=0$ contribution arises
through the 2$^-$ state ($j_x=2$).  In the case of $^{11}$Li, since
 $^{10}$Li shows a very
low-lying $2^-$ virtual state, the $^{11}$Li-resonances with a large
$2^-$-component ($3/2^+$ and $5/2^+$) will show a tendency to keep the
neutron close to the core,  leading therefore to a system smaller
than the $1/2^+$-state, where the $1^-$-components dominates.  This is
actually seen in the upper part of  Table~\ref{tabr2}, where the rms radii for the
$1/2^+$-state are significantly larger than those of the $3/2^+$ and
$5/2^+$-resonances.

In the case of $^{11}$O, the $2^-$-state in $^{10}$N is not that low
anymore, and it is actually rather broad (see Table~\ref{tab1}), being
then close to disappear into the continuum.  The confining effect of
the $s$-wave $2^-$-resonance disappears, and the $1/2^+$, $3/2^+$, and
$5/2^+$-resonances in $^{11}$O have a similar size, see
Table~\ref{tabr2}.

\subsection{Structure}

\begin{figure}
\begin{center}
\includegraphics[width=\columnwidth]{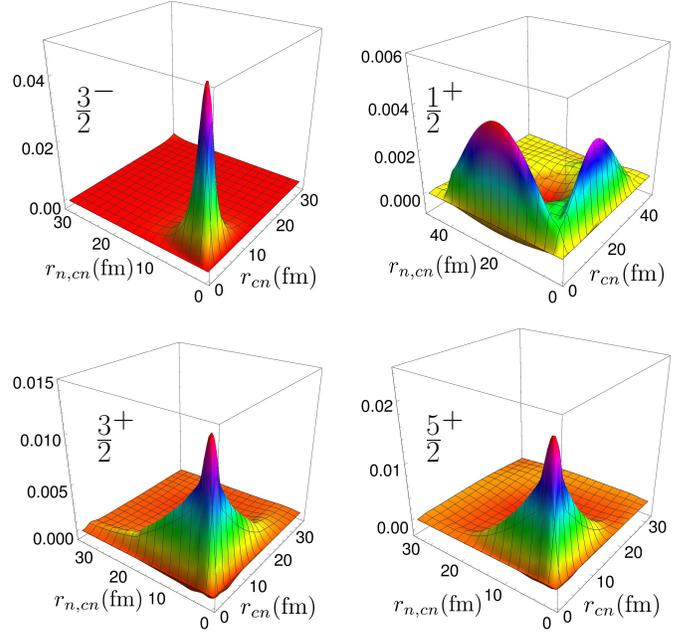}
\end{center}
\caption{Real part of the structure function $F(r_{cn},r_{n,cn})$, as defined in Eq.(\ref{funF}), in fm$^{-2}$, for the four computed states in $^{11}$Li.
A complex scaling angle $\theta=0.30$ rads has been used.}
\label{fig0}
\end{figure}

\begin{figure}
\begin{center}
\includegraphics[width=\columnwidth]{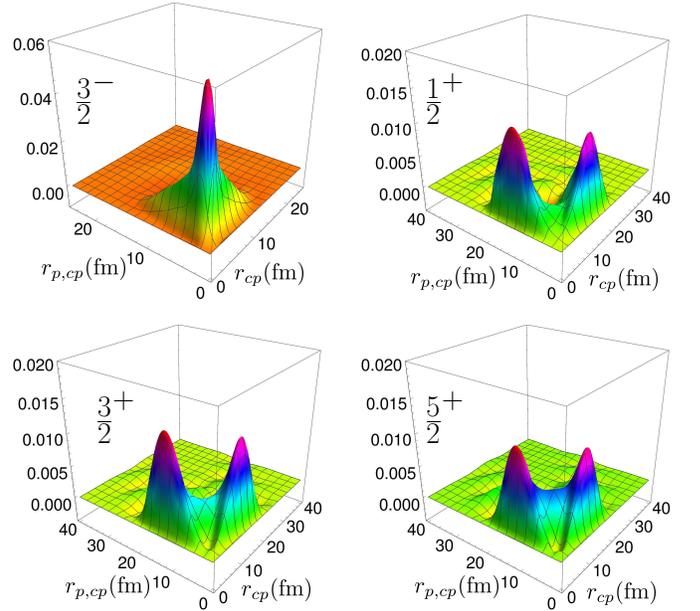}
\end{center}
\caption{The same as Fig.~\ref{fig0} for the $^{11}$O states.}
\label{fig4}
\end{figure}

The origin of the average distance results discussed above can be
visualized by means of the structure function
\begin{eqnarray}
\lefteqn{ \hspace*{-7mm}
F(r_{cN},r_{N,cN})=} \nonumber \\ & &
r_{cN}^2 r_{N,cN}^2 \int \left(\Psi(\bm{r}_{cN},\bm{r}_{N,cN})\right)^2 d\Omega_{cN} d\Omega_{N,cN},
\label{funF}
\end{eqnarray}
where $N$ represents either the neutron for $^{11}$Li or the proton
for $^{11}$O, $\Psi$ is the complex rotated three-body wave function
of a given $J^\pi$-state, and $\Omega_{cN}$ and $\Omega_{N,cN}$ are
the angles defining the directions of $\bm{r}_{cN}$ and
$\bm{r}_{N,cN}$, respectively.  Note that following the normalization
criteria described in Ref.\cite{moi98}, the definition of the
structure function above is made in terms of the square of the wave
function, and not in terms of the square of the modulus of the wave
function.  In principle, the function $F$ depends on the complex
scaling angle, but since $\Psi$ is normalized to 1 it is obvious that
$F$ satisfies that
\begin{equation}
\int F(r_{cN},r_{N,cN}) dr_{cN} dr_{N,cN}=1,
\label{funF1}
\end{equation}
which implies that the integral of the imaginary part is equal to zero.

The real part of the structure function $F$ is shown in
Figs.~\ref{fig0} and \ref{fig4} for all the computed states in
$^{11}$Li and $^{11}$O, respectively. For the resonances the complex
scaling angle $\theta=0.3$ rads has been used.  

For $^{11}$Li, the $3/2^-$ ground-state wave function is rather
confined, with a high peak centered around average distances
determined by $r_{cn}\approx r_{n,cn}\approx 6$ fm, as expected from
the rms radii shown in Table~\ref{tabr2} for this state.  When looking
at the 1/2$^+$, $3/2^+$, and $5/2^+$-resonances, we can see that the
wave function is progressively developing a tail along the $r_{cn}$-
and $r_{n,cn}$-axis, which in turn can be related to the lack of
barrier in the $s$-wave potential.  The main difference in the
structure function for these three states, is the presence of a peak
at relatively small core-neutron distances for the $3/2^+$ and
$5/2^+$-states. As explained above this is attributed to the important
contribution of the $2^-$-states, which present a very low-lying
virtual state. In the $1/2^+$-state this virtual state does not
contribute, the peak then disappears, and the wave function shows
mainly two wide peaks each located along the two axes.

For $^{11}$O, we see in Fig.~\ref{fig4} that the $3/2^-$-resonance
shows a structure function with a peak apparently similar to the one
of the $3/2^+$ and $5/2^+$-states, although in this case the peak is
essentially only of $p$-wave character.  For the $1/2^+$, $3/2^+$, and
$5/2^+$-resonances, the $s$-wave contributes significantly, but due
the Coulomb repulsion, which pushes up the $2^-$-resonance in
$^{10}$N, the $s$-wave potential is not able to keep the $s$-wave
proton sufficiently close to the core, and the peak observed for the
$3/2^+$ and $5/2^+$-states in $^{11}$Li disappears.

\section{Summary and conclusions}

We have calculated three-body properties of the four lowest excited
bound states or resonances for the two light mirror nuclei, $^{11}$O
and $^{11}$Li.  The phenomenological interactions are chosen to
reproduce all known properties of $^{11}$Li combined with
consistent information about the subsystem, $^{10}$Li.  The only
difference in interactions is that the Coulomb potentials are added in
$^{11}$O from the charges of the two protons and the $^{9}$C-core.  We
use the established hyperspherical adiabatic expansion method combined
with complex rotation to separate the resonances from the background
continuum structure.

The nuclei, $^{11}$O and $^{11}$Li, are special by being non-identical
mirrors on the neutron and proton driplines, that is at opposite sides
of the beta-stability curve.  The effect of the Coulomb interactions
is rather small for most nuclei, except for a substantial translation
of the absolute energies.  However, these smaller effects sometimes
carry signals about features of interest in a better understanding of
many-body nuclear structure.  In general the importance lies in change
of structure, which requires theoretical models beyond the mean-field.
A prominent example is the Thomas-Ehrman shift, but in general nuclei at
the driplines are expected to maximize such structure variation.

In this report, we predict the properties of $^{11}$O and $^{10}$N
from knowledge of $^{11}$Li and its two-body subsystem, $^{10}$Li.
Other investigations are available, but to our knowledge none consider
systematically all four lowest-lying excited states/resonances, and
their relation to the properties of the nucleon-core subsystems.
Furthermore, we use phenomenological interactions, which should
enhance the reliability of our predictions.  We emphasize that the
interactions in the present work are able to reproduce all known
features of $^{10}$Li and $^{11}$Li.

We first investigate the two-body nucleon-core subsystems,
proton-$^{9}$C and neutron-$^{9}$Li. In realistic calculations the
spin-spin splitting is essential, that is coupling of the $3/2^-$-core
and the $1/2^{\pm}$-proton angular momenta and parities.  The sequence
of the resulting states of $1^-$ and $2^-$ is not experimentally
determined. Fortunately, the only two crucial properties are, first the
degeneracy weighted centroid energy, and second that one of
these states is unbound with a marginally negative virtual energy.

The two-body potentials turn out to have attractive pockets at short
range for $2^-$, while overall repulsive for the $1^-$-state.  Both
receive additional repulsion from the Coulomb potentials in the
$^{10}$N-nucleus. By construction, the $2^-$-potential for $^{10}$Li
has a marginally unbound virtual state.  Both the $2^+$ and
$1^+$-potentials have attractive short-range pockets for both
$^{10}$Li and $^{10}$N.  The resonance energies of course follow the
pattern of the potentials with less than $0.6$~MeV for all $^{10}$Li
states and about $2$~MeV higher energies for $^{10}$N.

The computed three-body energy of $^{11}$Li is fine-tuned to reproduce
precisely the measured ground state value, while the three positive
parity excited states of both positions and widths are experimentally
unknown, but predicted to be very similar.  For $^{11}$O, we find in
contrast to $^{11}$Li that the $3/2^-$-state is about $1$~MeV higher
than the three positive-parity states.  The two lowest resonances, 1/2$^+$ 
and $3/2^+$, are very similar, and it is therefore as likely that one of
these is the ground state. This would be a qualitative difference
between important properties of these mirror nuclei.  This predicted
energy sequence in $^{11}$O is consistent with a perturbation estimate of the
Coulomb shift. However, it  is important to keep in mind that the uncertainty
introduced by the unknown three-body interaction, which is seen to 
play a more relevant role in the 3/2$^-$ state than in the positive-parity
states, could modify the ordering in the energy spectrum.

The structure of the wave functions is in principle revealed by the partial wave
decomposition.  It is striking that the positive parity resonances all
are of very similar partial wave content in $^{11}$Li and $^{11}$O.  In contrast,
the $3/2^-$-state in $^{11}$O is almost entirely made of $p$-waves of
both nucleon-core two-body states, whereas $p$-waves in $^{11}$Li only
contribute about $40\%$ and $s$-waves correspondingly by $57\%$. 

The spatial distributions of the nucleons surrounding the core also
differ substantially in the two nuclei.  For the $3/2^-$-state, the
two nucleons are symmetrically distributed in one peak in both cases,
but about $30\%$ closer to the core and more smeared out in the
$^{11}$O-resonance than in the $^{11}$Li bound-state.  The positive
parity resonances in $^{11}$O all three exhibit two peaks in their
density distributions corresponding to one proton close and one
further away from the core.  In $^{11}$Li, these two peaks coincide
for the $3/2^+$ and $5/2^+$-resonances, whereas they remain for
$1/2^+$, but with much larger tails. This fact shows that, even if
the partial wave content is similar (as shown in Table~\ref{tabcom2} for
the 3/2$^+$ and 5/2$^+$ resonances), the spatial distribution 
of the constituents can be different. 

In conclusion, all these detailed predictions are beyond present
laboratory tests, but still displaying essential properties, which in
turn should inspire to new experimental investigations.  The
substantial differences between the two mirror nuclei are all due to
the additional Coulomb interaction.  This is a new experience in
nuclear physics, where the Coulomb interaction generally is believed
to influence nuclear structure only marginally.  In summary, we have
learned that mirror nuclei not necessarily have very similar
structure.  Furthermore, we have seen that dripline nuclei still can
deliver new information about nuclear structure.

\acknowledgments We want to thank H.O.U. Fynbo and K. Riisager for
drawing attention to these systems and subsequent continuous discussions.
We also thank J. Casal for constructive comments and discussions.
This work has been partially supported by the Spanish
Ministry of Science, Innovation and University MCIU/AEI/FEDER,UE  (Spain) under Contract
No. FIS2018-093636-B-I00.

\end{document}